# Tailoring the MontiArcAutomaton Component & Connector ADL for Generative Development

Jan O. Ringert
School of Computer Science
Tel Aviv University
http://www.cs.tau.ac.il/

Bernhard Rumpe
Software Engineering
RWTH Aachen
http://www.se-rwth.de/

Andreas Wortmann
Software Engineering
RWTH Aachen
http://www.se-rwth.de/

## ABSTRACT

Component & connector (C&C) architecture description languages (ADLs) combine component-based software engineering and model-driven engineering to increase reuse and to abstract from implementation details. Applied to robotics application development, current C&C ADLs often require domain experts to provide component behavior descriptions as programming language artifacts or as models of a-priori fixed behavior modeling languages. They are limited to specific target platforms or require extensive handcrafting to transform platform-independent software architecture models into platform-specific implementations. We have developed the MontiArcAutomaton framework that combines structural extension of C&C concepts with integration of application-specific component behavior modeling languages, seamless transformation from logical into platform-specific software architectures, and a-posteriori black-box composition of code generators for different robotics platforms. This paper describes the roles and activities for tailoring MontiArcAutomaton to application-specific demands.

## 1. INTRODUCTION AND MOTIVATION

Engineering robotics software requires techniques to support comprehension, separation of concerns, and reuse of existing parts. Component-based software engineering (CBSE) is a technique to enable reuse of software components between applications [2]. Software components are defined as general-purpose programming language (GPL) artifacts. These are hardly reusable in different contexts (such as with different robots) and require domain experts to comprehend software engineering concepts as well as GPL details. Model-driven engineering (MDE) abstracts from this by lifting models to primary development artifacts. Models can be better comprehensible, platform-independent, and automatically translated into different implementations [8]. Component & connector (C&C) architecture description languages (ADLs) [18] combine CBSE and MDE to enable composition of complex software architectures from abstract component models. Code generators translate ADL models into executable systems.

Extensiblity is considered "a key property of modeling notations" [17] and crucial for development of applications for heterogeneous technologies as eminent in robotics. However, most C&C software ADLs come with fixed sets of modeling elements and component behavior modeling languages.

A recent survey on industrial use of ADLs found that existing extension mechanisms for tailoring an ADL "towards whole company needs" are insufficient and corresponding tools too generic [16]. This limits the principal design decisions the language may express and ultimately requires to realize important design decisions in GPL artifacts bundled with the architecture. Similarly, only 38% of the survey respondents generate code from ADL models. The most important reasons for handcrafting code are "no need", a "different level of abstraction between software architecture and target code", and "limited ADL expressiveness".

We present a methodology supported by the MontiArcAutomaton framework that combines structural extension of C&C concepts with integration of application-specific component behavior modeling languages, a transformation of platform-independent into platform-specific software architectures and composable code generators to complement language integration flexibility. With this,

R1 the C&C language can be extended to meet specific demands,

R2 domain experts describe component behavior using the most appropriate application-specific modeling languages or provide GPL implementations,

R3 logical software architectures are reused for different target platforms, and

R4 code generators are composed to translate architecture

J. O. Ringert acknowledges support from a postdoctoral Minerva Fellowship.

MORSE/VAO '15, July 21 2015, L'Aquila, Italy
c
DOI: http://dx.doi.org/10.1145/2802059.2802064

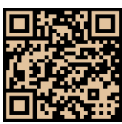

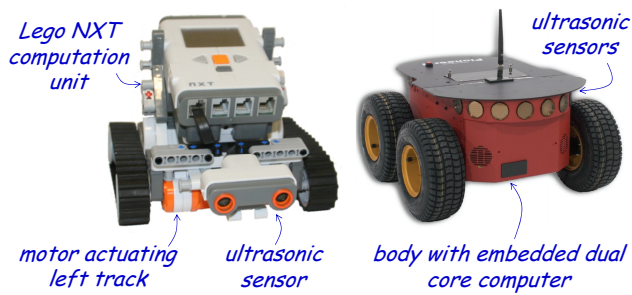

Figure 1: Exploration platforms: a Lego NXT robot with leJOS and a Pioneer 3-AT with ROS.

models of specific language combinations into implementations.

We have previously presented technical details on the extension mechanisms of MontiArcAutomaton in [27, 23, 15]. The description in this paper focuses on the methodology of applying these and comprises of roles and activities for the customization and extension of the MontiArcAutomaton framework and its ADL. In the following, Sect. 2 illustrates the issues MontiArcAutomaton tackles by example, before Sect. 3 presents background on MontiCore and MontiArcAutomaton. Sect. 4 describes how to use the extension mechanisms of MontiArcAutomaton and Sect. 5 discusses related work. Finally, Sect. 6 concludes.

## 2. EXAMPLE

Consider a robotics company that is going to produce robots for exploration of unknown areas. These robots are supposed to explore the area they are deployed to until they approach an obstacle, log it, back up, rotate, and start exploring again. The company is going to produce systems for different requirements: a cheap system for indoor education purposes and a robust system for outdoor exploration. Both systems are depicted in Fig. 1. The educational system - on the left - comprises of a Lego NXT robot equipped with an inexpensive computation unit, a front-mounted ultrasonic sensor, and the leJOS NXJ operating system[1] to interface the NXT hardware with Java. The outdoor exploration system uses Pioneer 3-AT robot with multiple front-mounted ultrasonic sensors, a powerful on-board computer, and the robot operating system ROS [20] to interface the hardware with Python.

Although both platforms require different GPLs, sensors, and actuators they provide similar functionality To reduce engineering costs by reusing the same logical software architecture for both systems (**R3**), the company uses an C&C ADL with embedded component behavior modeling languages and a GPL-independent data type language. A software engineer decomposes the system's functionality into nine component types as depicted in Fig. 2. The composed component type `ExplorerBot` contains two subcomponents of types `Button` and `UltraSonic` that provide input to a subcomponent of type `ExplorationControl`. `ExplorationControl` is composed from a `Controller`, which translates inputs into navigation commands, and a `Logger`, which persists distance data. The navigation commands are passed to a component of type `Navigation` that contains a `Translator` and two instances of component type `Motor` to translate navigation commands into control of two parallel motors that propel the robot.

[1]Website of leJOS: http://www.lejos.org/

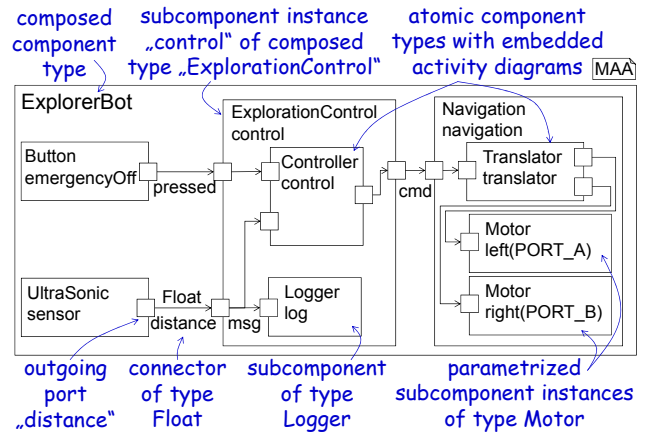

Figure 2: Platform-independent `ExplorerBot` Component and Connector model.

To ensure optimal component scheduling, the software engineer wants to specify their order in the `ExplorerBot` component type – a feature that requires to extend the ADL and add modeling elements to represent subcomponent schedules as well as well-formedness rules to check whether a schedule is valid (**R1**). The component types `Button`, `UltraSonic`, `Logger`, and `Motor` require specific GPL implementations as they use APIs to interface drivers of hardware components or operating system functionality. The integration and binding of specific implementations to components is a necessity for the project's engineers (**R2**). The engineers decide to model the behavior of the component types `Timer` and `Controller` with variant of UML activity diagrams (AD) popular at the company. Using ADs to model component behavior requires extending the ADL with the AD language and integration of new code generators (**R2** and **R4**).

## 3. PRELIMINARIES

In ongoing efforts we are creating the extensible C&C ADL MontiArcAutomaton[2] that integrates application-specific component behavior modeling languages with full support for extension and code generation to multiple platforms. Its powerful modeling language extension and integration mechanisms are implemented on top of the MontiCore language workbench, which we present in Sect. 3.1. We introduce the MontiArcAutomaton framework in Sect. 3.2.

### 3.1 The MontiCore Language Workbench

MontiCore [14] is a workbench for the compositional development of domain specific languages (DSLs). It comprises of a language to describe concrete and abstract syntax of DSLs and generators to create language processing infrastructure. For a specific DSL, tools to parse textual models and to translate these into an abstract representation, as well as frameworks for well-formedness checking [31], language integration [9], and template-based code generation [29] are generated from the DSLs context-free grammar.

[2]MontiArcAutomaton website: http://monticore.de/robotics/montiarcautomaton/.

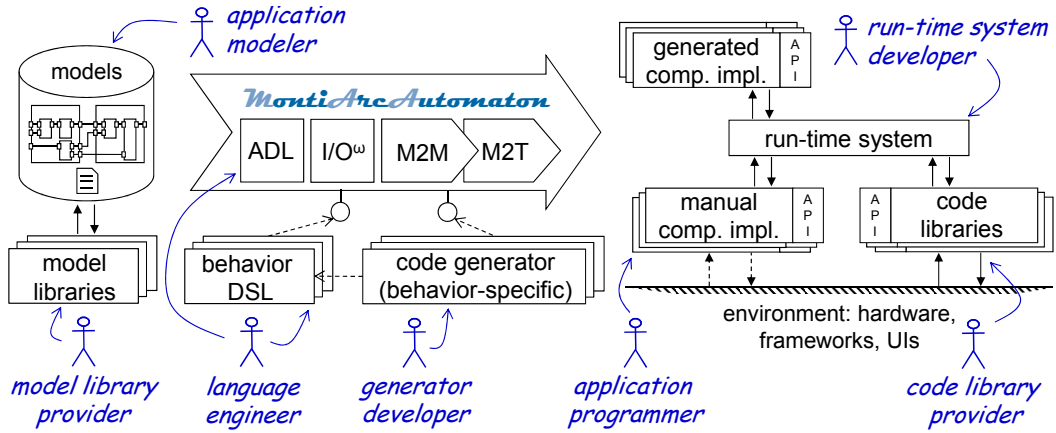

Figure 3: The MontiArcAutomaton framework processes C&C models with embedded component behavior models of application-specific languages to produce artifacts for arbitrary GPLs. To this end, it provides extension points for behavior languages and code generators that need to be implemented by different roles.

The language integration mechanisms of MontiCore realize a syntax-oriented approach towards black-box language integration of grammar-based languages. With these, MontiCore enables language aggregation, language embedding, and language inheritance. Language aggregation combines several languages into a loosely coupled *language family*, which enables a combined interpretation of independent models that reside in different artifacts. A typical application are orthogonal system aspects, e.g., a component modeling language and its data types. Language embedding combines languages by embedding elements of one language into distinguished extension points of another. This enables using both languages in a single artifact and lends itself to embedding component behavior modeling languages into component modeling language elements of ADLs. Language inheritance is used to refine or to extend an existing language. Inheritance is used to extend the MontiArcAutomaton ADL with additional modeling elements. In each case of reuse, the language processing infrastructure for all languages is generated by MontiCore. Its language integration mechanisms support all levels of syntax-oriented language integration, i.e., they integrate concrete syntax, abstract syntax, and well-formedness rules [31]. Detailed descriptions of MontiCore [13], language integration [10], and code generation [29] are available.

### 3.2 The MontiArcAutomaton Architecture Modeling Framework

At is core, MontiArcAutomaton contains an ADL extending the MontiArc ADL [11], embeds a component behavior modeling language based on $I/O^\omega$ automata [26], and aggregates UML/P class diagrams (CD) [28] to describe data types. All languages are implemented with MontiCore. The MontiArcAutomaton ADL [26] describes logically distributed software architectures as C&C systems in which components perform computations and connectors regulate communication. Components are black-boxes with stable interfaces of typed, directed ports and are either atomic or composed. The data types of ports are defined as UML/P CDs. Atomic components feature a component behavior description, either as a model of an embedded language [25], or as GPL artifacts. Composed components contain a hierarchy of subcomponents and their behavior emerges from their combination. MontiArcAutomaton distinguishes component types and subcomponent instances, supports generic types, component configuration parameters, and introduces component variables. Architecture models are parsed by the generated language processing infrastructure, checked for well-formedness, and transformed into executable system using multiple code generators [23, 24]. Different target platforms are supported by model and code libraries [27].

## 4. TAILORING MONTIARCAUTOMATON

Each tailoring activity is performed by roles involved in the MontiArcAutomaton development process (illustrated with the related artifacts in Fig. 3). We distinguish between the following roles and their specific technical skills:

*application modeler:* models C&C architecture and behavior, selects libraries, binds abstract to platforms-specific components

*application programmer:* provides GPL implementations of components if not possible to generate

*model library provider:* provides platform-independent components and data types

*code library provider:* implements modeled library component types in GPL if not possible to generate

*language engineer:* develops and integrate modeling languages

*generator developer:* provides code generators for languages and specific run-time system (RTS)

*RTS developer:* provides run-time system in platform GPL

The activities for tailoring the framework are organized in three stages addressing customization of the modeling language, development of the application model, and composition of code generators. An overview of the tailoring activities is shown in Fig. 4. Each stage starts with a decision node and has two alternative actions depending on the necessity of customization. The first stage distinguishes whether the MontiArcAutomaton language family has to be extended or not. The case of extending the MontiArcAutomaton ADL with new structural modeling elements and adding component behavior modeling languages is described in Sect. 4.1. In the second stage, a robotics application is modeled. Here either a platform specific architecture is developed or the architecture is kept platform-independent by means of model libraries and bindings. Details of the activity for developing a platform-independent architecture are provided in Sect. 4.2. Finally, in the third stage code generators are selected. If a monolithic code generator for the combination of ADL features, behavior languages, and target platform exists, it can be used. Otherwise, an appropriate code generator has to be composed from other code generators as described in Sect. 4.3. Engineering C&C applications with MontiArcAutomaton using its provided features, languages, and monolithic code generators, i.e., always selecting the upper branches in the activity shown in Fig. 4, are documented in [25, 24].

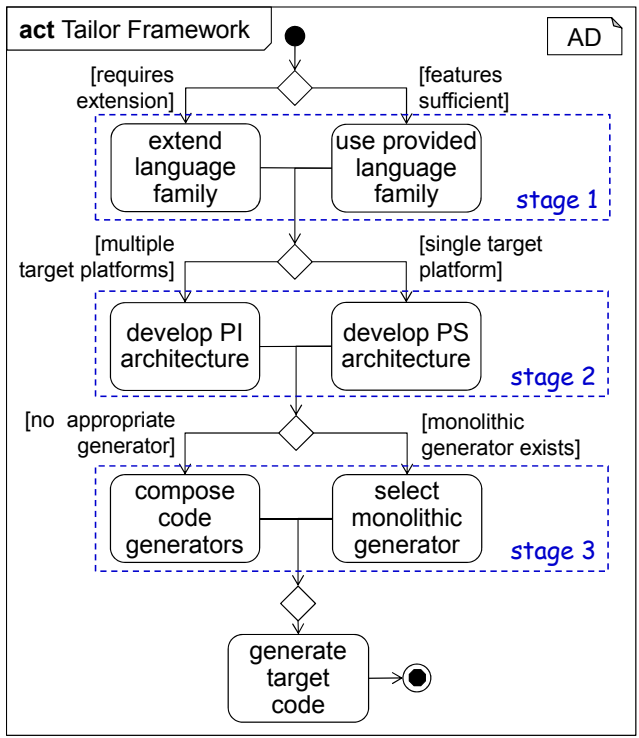

**Figure 4:** The stages and activities involved in tailoring MontiArcAutomaton, developing C&C applications, and generating platform-specific target code.

### 4.1 Language Family Extension

In case an existing MontiArcAutomaton language family is insufficient for the development of a new robotics application, it has to be extended. The extension mechanism comprises of extending the MontiArcAutomaton ADL (**R1**) and adding the most-appropriate component behavior modeling languages (**R2**). This activity is carried out by the roles *application modeler* and *language engineer* and shown in Fig. 5.

The *application modeler* examines the MontiArcAutomaton ADL for required features. These requirements are passed to the *language engineer*, who thus refines or extends the C&C modeling part of the MontiArcAutomaton ADL using the language inheritance feature of MontiCore. An example for this kind of extension is adding scheduling information as discussed in the example in Sect. 2 as **R1**. Afterwards, the application modeler identifies the required component behavior modeling languages. If new behavior languages have to be created or existing ones have to be extended, this task is passed to the language engineer again. In our running example an existing language for activity diagrams was chosen to extend the language family (see Sect. 2, **R2**). After all languages meet the application modeler's requirements, the language engineer integrates these into a new language family. In our example, this language family is also extended with well-formedness rules for scheduling information and embedded ADs.

### 4.2 Platform-Independent and Platform-Specific Architectures

The integration of components with platform-specific implementations renders an architecture platform-specific. When developing robotics applications for multiple platforms this dependency should be avoided (**R3**). MontiArcAutomaton uses platform-independent *model libraries*, platform-specific *code libraries*, and *bindings* to enable a late commitment to actual platforms [27]. We show how to apply these mechanisms for stage 2 in the activity in Fig. 6.

First, the application modeler identifies component types requiring platform-specific implementations that cannot be generated from behavior models. In our example scenario, the component types Button, UltraSonic, Logger, and Motor require specific GPL implementations because they interact with platform-specific APIs. The model library provider creates models of these components, which the application modeler then uses to model the platform-independent architecture. Platform-independent components that require GPL implementations are bound in a platform-specific application model. For each platform the application modeler identifies matching code libraries provided by the code library developer. In Sect. 2, e.g., platform-specific implementations of the UltraSonic sensor are required for the Lego NXT and Pioneer 3-AT robot.

### 4.3 Code Generator Development and Configuration

MontiArcAutomaton enables to use monolithic code generators for specific language aggregates, as well as composition of modular and reusable code generators (**R4**). The composition of modular code generators is aligned with the language extension mechanisms of the framework [23] and defines code generator kinds for three different concerns: (1) *component generators* produce code representing component hulls [30], i.e., ports, variables, messaging infrastructure, and the hierarchies of composed components, (2) *behavior generators* process models of a single embedded behavior language to produce component behavior code, and (3) *type generators* produce code to represent data types in the target GPL.

The activity of code generator composition begins with the *application modeler* identifying required code generators suitable for the intended target platforms. First, a proper *component generator* has to be selected. If no such generator exists, a *generator developer* needs to provide it. To integrate generated component code with handcrafted and generated behavior code, the component generator relies on a *run-time system* (RTS) provided by a *run-time system developer*. Afterwards, for each component behavior language, target platform, and RTS, a suitable *behavior generator* is selected. This may include the *generator developer* to develop proper generators. Finally, data type generators (e.g., for types used by ports) for the target platform's GPL are as well provided by a *code generator developer*. After selection of proper code generators, transformation into executable systems can be invoked.

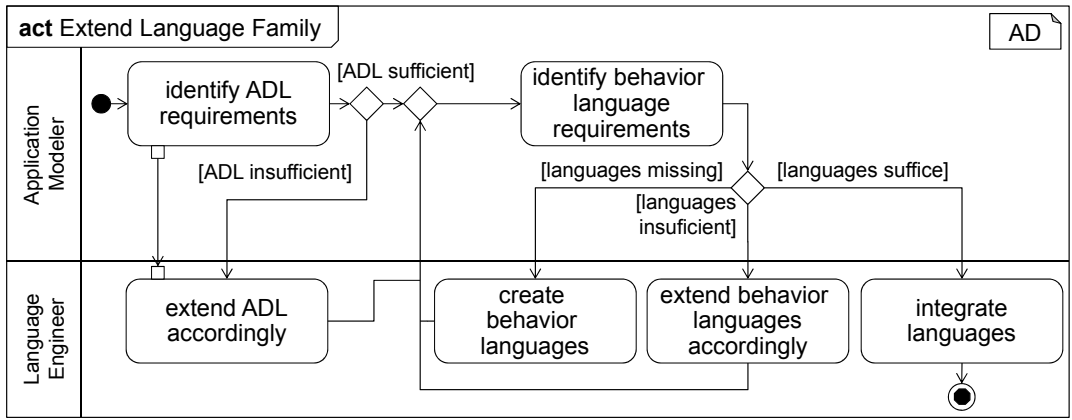

Figure 5: Extension and configuration of the MontiArcAutomaton language family.

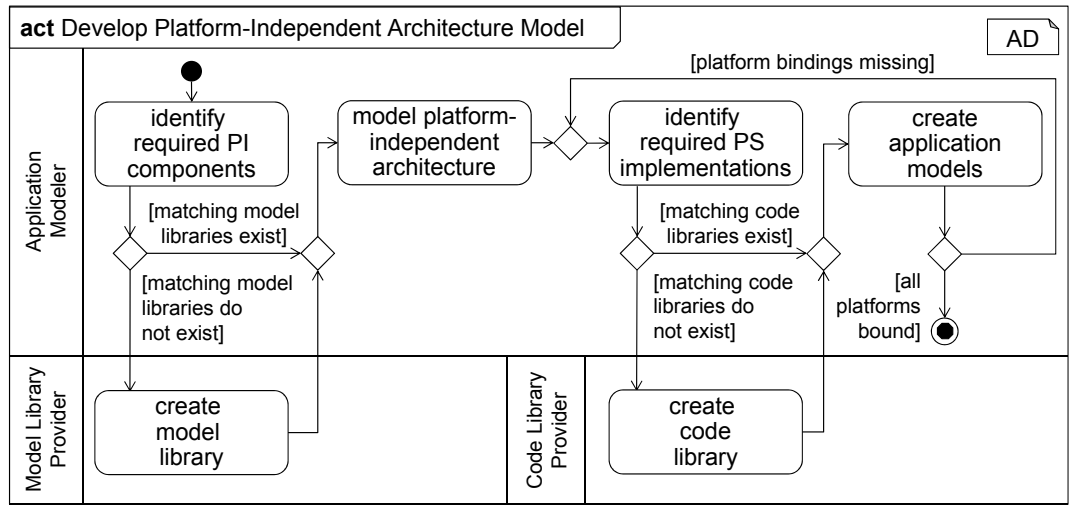

Figure 6: Developing a platform-independent software architecture and binding it to platform-specific implementations.

We now focus on the activity of developing new code generators shown in Fig. 7. This activity combines extension of the MontiArcAutomaton ADL, integration of new component behavior languages, introduction of a new data type language, and usage of a new target platform. The generator developer begins with identifying the missing generator's type. Component generators and behavior generators require selection of a matching RTS. Afterwards, the generator developer defines the generator's provided behavior, i.e., the processable component behavior languages. Then, she implements the generator's model-to-model or model-to-text transformations, and identifies the code generator's entry point and input requirements (cf. [23]). In collaboration with the language engineer, the generator developer identifies whether the code generator requires additional well-formedness rules to reject models using language features the code generator does not support (such as non-determinism). If such well-formedness rules are required, the generator developer creates these. Ultimately, the generator developer creates a generator model documenting the RTS, provided behaviors, entry point, and well-formedness rules of the new generator. MontiArcAutomaton relies on these models for actual generator composition as explained in [23].

## 5. RELATED WORK

Multiple architecture modeling languages and frameworks for C&C systems have been brought forth [18, 16, 21]. These have emerged from different domains, such as automotive [1, 12], avionics [7], or robotics [30, 6, 3] and focus different challenges of architecture engineering from academic and industrial perspectives. Most of these are "first-generation ADLs" [17] that are "solely focused" on technology instead of business-related or domain-specific aspects. The flexible integration of DSLs with ADLs is rare and usually overly com-plicated (cf. the "Behavior Annex" of AADL [7]). The ADL xADL [4, 5] focuses on architecture extensibility as well. It is based on the xArch[3] meta model for XML-based ADLs, shares many features with MontiArcAutomaton (such as atomic and composed component types, instantiation, component behavior models). Extension in xADL focuses on language extension on the meta model level and does neither support black-box language integration, nor integration of language processing infrastructure [19]. Also, xADL does not consider code generator composition. Instead, its architecture instantiation schemas [4] tie software architecture models to specific implementations. In architecture modeling frameworks driven by the demands of a specific domain, extensibility usually is less focused and domain-specific issues are challenged foremost. Popular robotics architecture modeling frameworks [21], such as SmartSoft [30], RobotML [6], BRICS [3], or SafeRobots [22], enable seamless development of robotics software architectures. These frameworks provide solutions to domain-specific issues, such as advanced communication patterns, deployment, or planning, that are not tackled by MontiArcAutomaton. Although most of these frameworks employ state of the art language workbenches, they neither focus ADL extension, nor integration of application-specific behavior languages, or code generator composition. The authors of [30] explicate this as "freedom from choice" to support application developers in creating solutions instead of dealing with framework mechanisms. To the best of our knowledge, neither generic ADLs, nor domain-specific ADLs provide an explicit extension methodology from language extension to generator composition.

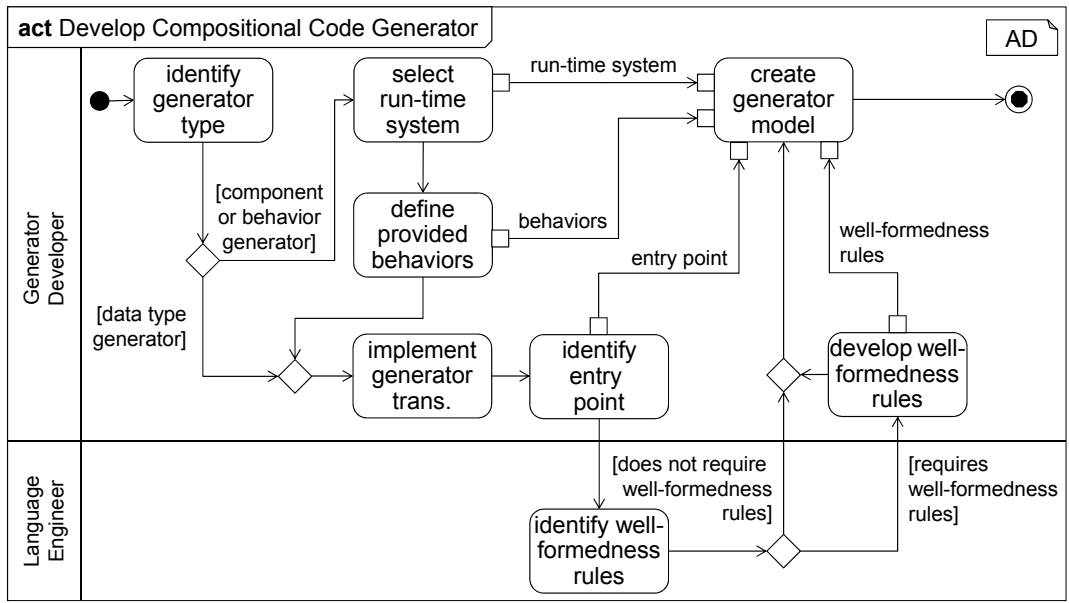

Figure 7: Activities to develop of a composable code generator.

## 6. SUMMARY

We have presented main activities for extending the ADL and framework MontiArcAutomaton. Our methodology combines structural extension of C&C concepts, integration of application-specific component behavior modeling languages, transformation of platform-independent into platform-specific software architectures and composable code generators to reflect language integration flexibility. The activities realizing this methodology rely on the language integration mechanisms of MontiCore [10], as well as on the C&C nature [26], binding concepts [27], and code generators [23] of MontiArcAutomaton. The presented methodology comprises of three stages: customization of the modeling language, development of the application models, and composition of code generators. Depending on the project's needs not all activities are necessary: off-the-shelf MontiArcAutomaton embeds I/O$^\omega$ automata as component behavior modeling language, can be used for modeling platform-specific software architectures, and provides code generators for Java, Python, and Mona [24]. With the presented methodology, ADLs can be tailored to specific needs and the resulting software architectures can be used with different platforms using platform-independent components and specifically composed code generators. In the future, we plan to examine how language engineers can be further assisted in creating ADL extension and valid language aggregates.

---

[3]xArch website: http://isr.uci.edu/architecture/xarch/.